\documentclass{JHEP3}

\newcommand{\ra}{{\rightarrow}}
\newcommand{\Tr}{{\rm Tr}}
\newcommand{\cN}{{\mathcal N}}
\newcommand{\CC}{{\mathbb C}}
\newcommand{\ZZ}{{\mathbb Z}}
\newcommand{\RR}{{\mathbb R}}
\newcommand{\tsigma}{{\tilde\sigma}}
\newcommand{\hsigma}{{\hat\sigma}}

\usepackage{multirow,epsfig,amsmath}
\title{Nonperturbative Tests of Three-Dimensional Dualities}
\author{Anton Kapustin\\ California Institute of Technology \\ Email: \email{kapustin@theory.caltech.edu}}
\author{Brian Willett\\ California Institute of Technology \\ Email: \email{bwillett@caltech.edu}}
\author{Itamar Yaakov\\ California Institute of Technology \\ Email: \email{itamar.yaakov@caltech.edu}}

\abstract{We test several conjectural dualities between strongly coupled superconformal field theories in three dimensions by computing their exact partition functions on a three sphere as a function of Fayet-Iliopoulos and mass parameters. The calculation is carried out using localization of the path integral and the matrix model previously derived for superconformal $\mathcal{N}=2$ gauge theories. We verify that the partition functions of quiver theories related by mirror symmetry agree provided the mass parameters and the Fayet-Iliopoulos parameters are exchanged, as predicted. We carry out a similar calculation for the mirror of $\mathcal{N}=8$ super-Yang-Mills theory and show that its partition function agrees with that of the ABJM theory at unit Chern-Simons level. This provides a nonperturbative test of the conjectural equivalence of the two theories in the conformal limit.
}
\keywords{Supersymmetric gauge theory, Chern-Simons Theories, Extended Supersymmetry, Matrix Models}
\preprint{68-2782}

\begin{document}

\section{Introduction}

Duality in interacting quantum field theories is a fascinating phenomenon. Supersymmetric gauge theories have provided us with many interesting examples of duality. One interesting class of examples is provided by mirror symmetry in three dimensions \cite{Intriligator:1996ex}. This duality relates superconformal field theories arising as the IR fixed points  of three dimensional supersymmetric quiver gauge theories.  The moduli space and global symmetries of these dual pairs were compared in \cite{Intriligator:1996ex,HW,deBoer:1996mp,deBoer:1996ck}. The matching of moduli spaces provided the original motivation for the conjecture. Duality, in this case, exchanges the Higgs and Coulomb branches of the moduli space, as is the case with four dimensional gauge theories arising from compactification of string theory on mirror Calabi-Yau manifolds. The Abelian subgroup of the flavor symmetries on one side maps to the topological symmetries associated with $U(1)$ factors of the gauge group on the other. Real mass parameters for the hypermultiplets are exchanged with Fayet-Iliopoulos (FI) parameters for the vector multiplets. The full flavor symmetry group is realized non-locally in the dual theory and is not visible at the level of the action. We will explore only the $\mathcal{N}=4$ version of the duality and utilize the map of symmetries and parameters given in \cite{deBoer:1996ck}.    

In general, it is very difficult to establish dualities, or gather evidence for their existence, when one or more of the theories involved is strongly coupled. In the case of mirror symmetry in three dimensions, the duality applies strictly only to the IR limit of the gauge theories involved, a limit in which the gauge coupling runs to infinity. A perturbative comparison of quantities on the two sides of the duality is therefore not possible. One may still hope to compare quantities and features which do not depend on the gauge coupling. The moduli space of the theory is one such feature. Another is the expectation value of supersymmetric observables, such as the partition function  regarded as a function of the FI and mass parameters. The phenomenon of localization of the path integral makes the computation of such quantities feasible. The expectation value of an observable which preserves some fraction of the supersymmetry of the theory will, generically, receive contributions only from a very limited subset of the space of fields involved in the path integration. In the most radical case, this reduced space is simply a set of points, as happens for chiral operators of some simple $\mathcal{N}=(2,2)$ sigma models in two dimensions. A more interesting situation arises when the path integral reduces to a finite dimensional integral, as in the case of matrix models. In general, one can have a very non-trivial instanton moduli space as the domain of integration, but this does not happen in the present context.  

In this paper we provide evidence for mirror symmetry in three dimensions by comparing the partition functions of a class of dual theories on a three-sphere. We limit ourselves to theories with unitary gauge groups, since only in this case one can introduce the FI parameters.\footnote{If one does not introduce the FI deformations and their dual mass deformations, the partition function is a number which depends on the details of the normalization of the measure in the path-integral.} Our calculations rely heavily on the derivation of a class of matrix models, described in \cite{Kapustin:2009kz}, which is appropriate for a large class of $\mathcal{N}=2$ superconformal gauge theories in three dimensions. The IR limits of the $\mathcal{N}=4$ quiver gauge theories involved in mirror symmetry fall into this category. There are many supersymmetric observables in these theories, some of which may have a non-trivial expectation value, but we restrict ourselves to the partition function. It is worth noting that the dual observables are not known, in general.

As another application of the localization method, we consider $\mathcal{N}=8$ SYM theory in three dimensions. This theory arises as the low energy effective theory on coincident D2 branes in type IIA string theory. In the case when the gauge group is $U(N)$, its IR fixed point describes the dynamics of $N$ coincident M2 branes in M theory. Recently, Aharony, Bergman, Jafferis, and Maldacena proposed an action to describe this system \cite{Aharony:2008ug}. The action is of Chern-Simons type, with a product gauge group and opposite CS levels for the two factors. It describes coincident M2 branes probing a $\CC^4/\ZZ_k$ orbifold, where $k$ is the Chern-Simons level. The ABJM theory is believed to be holographically dual to M theory on $AdS_{4}\times S^{7}/\ZZ_k$. For general $k$ it has $\mathcal{N}=6$ superconformal symmetry, but has been conjectured to have $\mathcal{N}=8$ supersymmetry when $k=1$ or $k=2$ and to be isomorphic to the IR fixed point of the $\cN=8$ SYM theory for $k=1$ \cite{Aharony:2008ug,Benna:2009xd}. It is rather hard to test this conjecture, because both theories are strongly coupled. We compute the partition functions for these two theories deformed by FI and mass parameters and find agreement. This provides a nonperturbative test of the conjecture.\footnote{A similar test of this duality involving the partition function on $S^2 \times S^1$ was performed earlier by S. Kim.\cite{Kim:2009wb}}

The outline of the paper is as follows. In section \ref{Setup}, we review the necessary background material. We review $\cN=4$ $d=3$ supersymmetric gauge theories, briefly explain the idea behind localization, write down a matrix model appropriate for $\mathcal{N}=4$ gauge theories, and describe the theories which are conjectured to be related by mirror symmetry. In section \ref{Quivers}, we compare the partition functions of mirror pairs of theories and show that they agree provided the FI and mass parameters are matched as proposed in  \cite{Intriligator:1996ex,HW,deBoer:1996mp,deBoer:1996ck}. In section \ref{ABJM}, we perform the calculation of the partition function for the special case of $\mathcal{N}=8$ SYM, which has some special features, and compare with the theory of ABJM  \cite{Aharony:2008ug}.  Appendix \ref{Identity} contains the proof of an identity involving hyperbolic functions used in the body of the paper.

This work was supported in part by the DOE grant DE-FG02-92ER40701.

\section{\label{Setup}Ingredients and methods}

\subsection{Supersymmetric gauge theories in three dimensions}

The $\mathcal{N}=4$ supersymmetry algebra in three dimensions has 8 real supercharges (not counting possible conformal supercharges). It is convenient to work in $\mathcal{N}=2$ superspace, in which 4 real supersymmetries are realized off-shell. After Euclideanizing the theory, $\mathcal{N}=2$ supersymmetry will be realized with two complex two-component spinors. The maximal possible R-symmetry group of such theories is $Spin(4)_R=SU(2)_R\times SU(2)_N$, with supercharges transforming as the $(2,2)$ representation. Some of the multiplets involved in three dimensional mirror symmetry are summarized in Table \ref{multiplets}. Note that the content of these multiplets is not symmetric with respect to the exchange of $SU(2)_R$ and $SU(2)_N$, so along with the multiplets listed in the table there exist also ``twisted'' multiplets with the roles of $SU(2)_R$ and $SU(2)_N$ exchanged. We will denote twisted vector and hyper multiplets with a hat.

Our starting point is the action for the $\mathcal{N}=4$ quiver theories considered in \cite{deBoer:1996ck} comprising the following components:

\begin{itemize}
\item Standard kinetic terms and minimal gauge couplings for all matter fields.
\[{S_{matter}} =  - \int {d^3 x {d^2}\theta {d^2}\bar \theta } \sum\limits_{matter} {({\phi ^\dag }{e^{2V}}\phi }  + {{\tilde \phi }^\dag }{e^{ - 2V}}\tilde \phi )\]

\item A Yang-Mills term for each factor in the gauge group. The gauge couplings for the different factors need not be the same, but all flow to strong coupling in the IR.
\[{S_{gauge}} = \frac{1}
{{{g^2}}}\int {d^3 x {d^2}\theta {d^2}\bar \theta } (\frac{1}
{4}{\Sigma ^2} - {\Phi ^\dag }{e^{2V}}\Phi )\]
where $\Sigma$ is the linear multiplet which includes the field strength for the connection $A_{\mu}$, and $\Phi$ is the adjoint chiral multiplet that is part of an $\mathcal{N}=4$ vector multiplet.

\item The possible holomorphic superpotential of the $\mathcal{N}=2$ theory is restricted by $\mathcal{N}=4$ supersymmetry to take the form 
\[{S_{sp}} =  - i\sqrt 2 \int {d^3 x {d^2}\theta \sum\limits_{matter} {(\tilde \phi \Phi \phi ) + c.c} } \]
where the sum runs over all matter charged under the gauge symmetry associated with $\Phi$. The term couples the adjoint representation to the tensor product of the representations $R$ and $R^{*}$. 
\end{itemize} 

We also consider two possible deformations of the theory:

\begin{enumerate}
\item Real and complex mass terms for the hypermultiplets. These transform as a triplet of $SU(2)_N$ and can be viewed as the lowest components of a background $\mathcal{N}=4$ vector multiplet coupled to the flavor symmetry currents. The $\mathcal{N}=4$ coupling is
\[{S_{mass}} =  - \int {d^3 x {d^2}\theta {d^2}\bar \theta } \sum\limits_{matter} {({\phi ^\dag }{e^{2{V_m}}}\phi }  + {{\tilde \phi }^\dag }{e^{ - 2{V_m}}}\tilde \phi ) - \left(i\sqrt 2 \int {d^3 x {d^2}\theta \sum\limits_{matter} {(\tilde \phi {\Phi _m}\phi ) + c.c} }\right)\]

When we consider localization, the conditions for the vanishing of the fermion variations of the background multiplet will imply $V_{m}\propto m\bar{\theta}\theta$ and $\Phi_{m}=0$, where $m$ is the real mass parameter.
  
\item Fayet-Iliopoulos (FI) terms for the $U(1)$ factors of the gauge group. These transform as a triplet of $SU(2)_{R}$. They can be viewed as the lowest components of a background twisted $\mathcal{N}=4$ vector multiplet coupled to the topological currents associated with the $U(1)$ factors by a $BF$ type coupling. The $\mathcal{N}=4$ coupling is
\[{S_{FI}} = Tr\int {d^3 x {d^2}\theta {d^2}\bar \theta \Sigma {{\hat V}_{FI}}}  + Tr\left(\int {d^3 x {d^2}\theta \Phi {\hat\Phi _{FI}} + c.c} \right)\]
where the $Tr$ picks out the $U(1)$ factors of the gauge group. Localization will require  ${\hat V}_{FI}\propto \eta\bar{\theta}\theta,\hat\Phi_{FI}=0$.
\end{enumerate}

\begin{table}
\begin{tabular}{|l|l|c|c|c|c|l|}
  \hline
$\mathcal{N}=4$ & $\mathcal{N}=2$ & Components & $SU(2)_{E}$ & $U(1)_{R}$ & $SU(2)_{L}\times SU(2)_{R}$ & $G$ \\
\hline\hline
vector & vector & $A_{\mu}$ & $1$ & $0$ &  & \multirow{7}{*}{adjoint} \\ 
 multiplet & multiplet ($V$) & $\lambda_{\alpha}$ & $\frac{1}{2}$ & 1 & & \\
& & $\sigma$ & $0$& $0$ &  $\big \{ \sigma, Re\varphi, Im\varphi \big \}  \quad \big|\quad \left(0,1\right)$ &  \\
& &  $D$ & $0$ & $0$  & $\big \{\lambda_{\alpha},\xi_{\alpha} \big \}  \quad \big|\quad \left(\frac{1}{2},\frac{1}{2}\right)$ &  \\
 \cline{2-5}
 & chiral & $\varphi$ & $0$ & $0$ & $\big \{ D, ReF_{\Phi}, ImF_{\Phi} \big \}  \quad \big|\quad \left(1,0\right)$ & \\
 & multiplet ($\Phi$) & $\xi_{\alpha}$ & $\frac{1}{2}$ & $1$ & & \\
&  & $F_{\Phi}$ & $0$ & $0$ & & \\
 \cline{2-6} 
\hline
hyper & chiral & $\phi$ & $0$ & $0$ &  & \multirow{3}{*}{$R$} \\
 multiplet & multiplet ($\phi$) & $\psi_{\alpha}$ & $\frac{1}{2}$& $1$ & $\big \{\phi,\tilde{\phi}^{\dag}  \big \}  \quad \big|\quad \left(\frac{1}{2},0\right)$ & \\
&  & $F$ & $0$ & $0$ & $\big \{ \phi^{\dag},\tilde{\phi}  \big \}  \quad \big|\quad \left(0,\frac{1}{2}\right)$ & \\
 \cline{2-5} \cline{7-7}
 &  chiral &  $\tilde{\phi}$ & $0$ & $0$ &  $\big\{ \psi_{\alpha},\tilde{\psi}_{\alpha}  \big \}  \quad \big|\quad \left(\frac{1}{2},\frac{1}{2}\right)$ & \multirow{3}{*}{$R^{*}$} \\
 & multiplet ($\tilde{\phi}$) & $\tilde{\psi}_{\alpha}$ & $\frac{1}{2}$& $1$ &$F,\tilde{F} \quad \big|\quad$ integrated out & \\
&  & $\tilde{F}$ & $0$ & $0$ & & \\
\hline
\end{tabular}
\caption{Field content and charges of the supersymmetry multiplets involved in three dimensional mirror symmetry.}
\label{multiplets}
\end{table}

\subsection{The infrared limit}\label{sec:IR}

The theories described above are super-renormalizable and flow to a free theory in the ultraviolet. In the infrared they flow to an $\cN=4$ superconformal field theory. This theory is typically nontrivial when the expectation values of all fields and the FI and mass deformations are set to zero. The conformal dimensions of fields in the infrared limit are determined by their transformation properties with respect to the $Spin(4)$ R-symmetry which is part of the superconformal symmetry. Typically, this R-symmetry coincides with $SU(2)_R\times SU(2)_N$ symmetry which is manifest in the Lagrangian of the theory. If this is the case, then the scalars in the vector multiplets and the gauge field have the infrared conformal dimension $1$, while the scalars in the hypermultiplets have the infrared conformal dimension $1/2$. This implies that in the infrared limit the kinetic terms of the vector multiplets are irrelevant and may be dropped. In other words, in such theories the infrared limit is the limit $g\ra\infty$.

It may happen that part or all of the infrared $Spin(4)$ R-symmetry is ``accidental'', i.e. arises only in the infrared limit, and does not coincide with the $SU(2)_R\times SU(2)_N$ R-symmetry visible at a finite energy scale. One such case is the $\cN=8$ super-Yang-Mills theory which may be regarded as an $\cN=4$ gauge theory with a single adjoint hypermultiplet. This theory is believed to flow to an $\cN=8$ superconformal field theory. We may of course think of the infrared limit of this theory as an $\cN=4$ superconformal field theory. But the manifest $SU(2)_R\times SU(2)_N$ symmetry cannot be identified with the $Spin(4)$ R-symmetry of this superconformal field theory. This is most easily seen in the abelian case, where the $\cN=8$ superconformal field theory is free. In that case the free scalars scalars transform as a spinor of the $Spin(8)$ R-symmetry. Decomposing the spinor representation with respect to the obvious $Spin(4)$ subgroup, we see that they transform as a pair of spinors of $Spin(4)$ with opposite chirality. On the other hand, from the point of view of $SU(2)_R\times SU(2)_N$ these eight free fields transform as  $(1,1)\oplus (2,2)\oplus (1,3)$, i.e. as a singlet, a vector, and a self-dual tensor. (The singlet is the dual of the abelian gauge field). 

If the dimensions of the fields in the vector multiplet are such that their Yang-Mills kinetic terms cannot be dropped in the infrared, the localization method developed in \cite{Kapustin:2009kz} does not apply.  For this reason in this paper we will only discuss $\cN=4$ theories where the superconformal $Spin(4)$ symmetry coincides with the manifest $SU(2)_R\times SU(2)_N$ symmetry. This means of course that we cannot directly apply our method to the infrared limit of the $\cN=8$ super-Yang-Mills theory. We circumvent this difficulty by studying a mirror realization of the same $\cN=8$ superconformal field theory, see section 5 for details.

\subsection{Localization on a three-sphere and the matrix model}

Next we discuss the localization of the path-integral for the infrared limit of the $\cN=4$ $d=3$ gauge theories on $S^3$. It was shown in \cite{Kapustin:2009kz} that the path-integral localizes to a matrix model integral. The necessary input for constructing the latter is the gauge group and matter representations. The matter for $\mathcal{N}=4$ theories comes in pairs of conjugate representations, which results in a substantial simplification in the formulas \cite{Kapustin:2009kz}. 

As discussed above, the infrared limit for the class of theories we are considering is the strongly-coupled limit $g\ra\infty$. We will study the partition function of this infrared theory deformed by the FI and mass terms. Supersymmetry guarantees that no renormalization of the FI or mass parameters is required. Setting $g=\infty$ means dropping the kinetic terms for the ${\mathcal N}=4$ vector multiplets.  However, the $Q$-exact localizing term can be thought of as a regulator, giving these fields non-degenerate kinetic terms.

To localize the path-integral to a finite-dimensional submanifold in the infinite-dimensional space of field configurations we pick a single supercharge $Q$, corresponding to a Killing spinor $\epsilon$ on $S^{3}$, and add to the action a term which is $Q$ exact with an arbitrary coefficient $t$ 
$$
{S_{loc}} = t\, \int d^3 x \quad Q\left(\sum\limits_{fermions} {\{ Q,{\Psi _i}} {\} ^\dag }{\Psi _i}\right)
$$
where ``fermions'' refers to all fermions in the theory. The bosonic part of this term is positive semi-definite. The addition of such a term does not alter the expectation value of $Q$-closed observables (e.g. the partition function). In the limit as $t\rightarrow\infty$, the path integral calculation for such observables localizes to field configurations for which the bosonic part of the $Q$-exact term vanishes. In fact, in this limit the semiclassical approximation around the zero locus of the localizing term is exact. One need only consider the classical contribution of the action and a possible one-loop determinant. Both contributions were calculated in \cite{Kapustin:2009kz} and the results are summarized below.

\begin{itemize}
\item The field configurations for which the above $Q$ exact term vanishes have all matter fields set to zero. For the fields in the vector multiplets,  the vanishing of the $Q$ exact term implies that $\sigma=-D=\sigma_{0}$  is a constant on $S^3$ and all other fields are set to zero. The path integral reduces, therefore, to an integration over the Lie algebra of the gauge group parameterized by the constant $\sigma_{0}$. Gauge-invariance allows one to assume that $\sigma_0$ is in a Cartan subalgebra of the Lie algebra, at the expense of the introduction of a Vandermonde determinant into the integration measure (see below). All of this also applies to the background vector multiplets responsible for the mass and FI parameters; in particular, localization requires $\Phi_m=\hat\Phi_{FI}=0$. 

\item There are no classical contributions from the original action. The BF coupling for the background vector multiplet responsible for the FI terms contributes, for every $U(1)$ factor of the gauge group, a factor of

\[S^{classical}_{FI} =2\pi i \eta \Tr\, \sigma_0  \]
 
\item The one-loop determinant coming from fluctuations of the fields around the saddle points consists of two parts
\begin{itemize}
\item For every $\mathcal{N}=4$ vector multiplet there is a factor
\begin{equation}\label{eq:vect}
Z^{vector}_{1 - loop} = \prod\limits_\alpha  \Big( \frac{{\sinh (\pi \alpha (\sigma_0 ))}}{{\pi \alpha (\sigma_0 )}}\Big)^2
\end{equation}
where the product is over the roots of the Lie algebra of G.
\item For every $\mathcal{N}=4$ hypermultiplet (matter) there is a factor
\begin{equation}\label{eq:hyper}
Z^{hyper}_{1 - loop} = \prod\limits_\rho   \frac{1}{\cosh(\pi \rho (\sigma_0 ))}
\end{equation}

where the product is over the weights of the representation $R$. When a background vector multiplet generating a mass parameter is included, the effect is just a shift

\[Z^{hyper+background}_{1 - loop} = \prod\limits_\rho   \frac{1}{\cosh(\pi \rho (\sigma_0 +\omega))}\]

where $\omega$ is the $\bar{\theta}\theta$ component of the background vector multiplet (the real mass). That this is the only effect can be easily deduced from the couplings described in the previous section.

\end{itemize} 
\item In addition to these, gauge fixing the matrix model by choosing $\sigma_{0}$ to be in the Cartan of the Lie algebra of $G$ introduces the standard Vandermonde determinant
\[{\prod\limits_\alpha  {\alpha (\sigma_0 )} ^2}\]
where the product is over all roots of the Lie algebra of $G$.
This cancels nicely for every factor of the gauge group with the denominator of the term coming from the $\mathcal{N}=4$ vector multiplet.

\item Finally, we divide by the order of the Weyl group to account for the residual gauge symmetry remaining after gauge-fixing $\sigma_{0}$ to the Cartan subalgebra.

\end{itemize}

The contribution of a vector multiplet requires some comment. An $\cN=4$ vector multiplet consists of an $\cN=2$ vector multiplet and an $\cN=2$ chiral multiplet in the adjoint representation. According to \cite{Kapustin:2009kz} the contribution of the $\cN=2$ vector multiplet is given precisely by (\ref{eq:vect}). Thus we are claiming that the contribution of the $\cN=2$ chiral multiplet which is part of the $\cN=4$ vector multiplet is $1$.  This might seem surprising since according to eq. (\ref{eq:hyper})  the contribution of a similar chiral multiplet which is part of a hypermultiplet in the adjoint representation is not $1$. The reason for this difference is that the two kinds of chiral multiplets have different infrared conformal dimensions and therefore different transformation properties with respect to the fermionic symmetry $Q$. For example, the lowest (scalar) component of the chiral multiplet has infrared conformal dimension $1$ or $1/2$ depending on whether the chiral multiplet is part of an $\cN=4$ vector multiplet or a hypermultiplet. In principle, to determine the contribution of the chiral multiplet which is part of the $\cN=4$ vector multiplet we need to redo the localization computation of \cite{Kapustin:2009kz} using an appropriately modified formula for $Q$.  

However, it is simpler to note that we are free to add the following $Q$-exact term for the adjoint chiral $\Phi$:

\[ s \int d^3 x d^2 \theta \Phi^2 \]

Taking $s$ to be large and positive, $\Phi$ is localized to zero, and the corresponding $1$-loop determinant is unity.  By $R$-symmetry, such a term is only allowed for the chiral multiplet which appears in the $\mathcal{N}=4$ vector multiplet.

\subsection{Mirror symmetry in three dimensions}

We will apply the localization procedure described above to some $\mathcal{N}=4$ quiver theories, which are conjectured to be related by mirror symmetry \cite{Intriligator:1996ex,HW,deBoer:1996ck}. Mirror symmetry for such theories can be deduced by combining Hanany-Witten-type brane constructions in Type IIB string theory \cite{HW} and the $SL(2,\ZZ)$ duality symmetry. Details of the brane construction and the resulting interpretation of mirror symmetry in three dimensions can be found in \cite{HW, deBoer:1996ck}. We will freely use string theory terminology in what follows, even though all our computations are purely field-theoretic.

The quiver theories which we will analyze are specified by the following data

\begin{itemize}
\item The theory has a gauge group
\[G = {U(N)}^n \]
Every factor $U(N)$ is associated with a set of $N$ coincident D3 branes in Type IIB string theory. Branes associated to adjacent factors end on the same NS5 brane, of which there are $n$ in total.  The dimension along which the fivebranes are spaced is compactified to a circle, and so the first and last factors are considered adjacent. 

\item For every gauge group factor there are $v_{i}$ fundamental hypermultiplets, $v_i\geq 0$. These are associated with $v_i$ D5 branes intersecting the $i$'th set of D3 branes.
\item There is an additional bifundamental hypermultiplet for every adjacent pair of gauge group factors.  These come from fundamental strings crossing the NS5 branes.
\end{itemize}

In the mirror theory, the D5 and NS5 branes are exchanged.  The gauge group is
\[G = {U(N)}^v \]
where $v=\sum_i v_i$. 
For every $i$ there is a fundamental hypermultiplet associated to the $j$'th gauge group factor, where
\[j = \sum\limits_{l = 1}^{i - 1} {{v_l}} \]
and for the first factor we sum $l$ from $1$ to $n$. Note that some of the $v_{i}$'s may vanish, so two $i$'s may contribute a fundamental hypermultiplet to the same gauge group factor. As before, there is an additional bifundamental hypermultiplet for every adjacent pair of gauge group factors. 
The global symmetries of the dual theories, and the map between them, are described in section \ref{BackgroundFields}.

Note that if $v_i=0$ for all $i$, then the vector multiplet corresponding to the diagonal $U(1)$ subgroup in the gauge group is free. In the dual theory the gauge group in this case becomes $U(N)$, and the bifundamental  hypermultiplet becomes the adjoint hypermultiplet. Its trace part is also free. These two free fields are exchanged by mirror symmetry. We may simply drop them and obtain a mirror pair of theories without decoupled fields. This remark will prove useful since our method of computing the partition function does not apply to the theory of a free vector multiplet.

\section{Abelian mirror pairs}

In the case when all gauge group factors have rank one (i.e., $N=1$), the statement of mirror symmetry formally follows from a simpler statement: $U(1)$ gauge theory with a single charged hypermultiplet is mirror to a free hypermultiplet \cite{Kapustin:1999ha}. Under this correspondence the FI parameter of the $U(1)$ gauge theory maps to the mass parameter of the free hypermultiplet. 

Let us verify that the partition functions of these two theories agree. According to the rules formulated above, the partition function of a free hypermultiplet of mass $\omega$ is
$$
\frac{1}{\cosh\pi\omega}.
$$
The partition function of $U(1)$ gauge theory with an FI term and a single charge-1 hypermultiplet is
$$
\int d\sigma \frac{e^{2\pi i\sigma\eta}}{\cosh\pi\sigma}.
$$
Computing this integral using residues we find that the integral evaluates to $1/\cosh(\pi\eta)$. Thus the partition functions indeed agree provided we identify $\eta$ and $\omega$. 

We see that this test of basic abelian mirror symmetry goes through because the partition function of a free hypermultiplet regarded as a function of mass coincides with its own Fourier transform. On the other hand, according to \cite{Kapustin:1999ha} basic abelian mirror symmetry is essentially equivalent to the statement that the partition function of a free hypermultiplet, regarded as a functional of a background $U(1)$ vector multiplet, coincides its own ``functional Fourier transform''. Since a real mass term $\omega$ can be regarded as a background vector multiplet of the form $V\sim \omega\bar\theta\theta$, the two statement are clearly very similar. Localization implies that the former statement follows from the latter. Of course, it is much harder to verify the functional Fourier transform property for a general value of the background vector multiplet because it seems impossibly difficult to evaluate the partition function of a free hypermultiplet in an arbitrary background. To prove the basic abelian mirror symmetry one needs to use less direct methods \cite{Borokhov:2002cg}. 

As in \cite{Kapustin:1999ha}, the fact that the partition functions match for the basic abelian mirror symmetry implies that they match for all abelian mirror pairs. As an example, let us show that the test goes through for the mirror pair of type $A_{n-1}$ considered in \cite{Intriligator:1996ex}. The first theory in this pair is a quiver theory with $N=1$ and $v_i=0$ for $i=1,\ldots,n$. The second theory has gauge group $U(1)$, $n$ hypermultiplets of charge $1$, and a decoupled free hypermultiplet (the adjoint of $U(1)$). Recalling the remark at the end of the previous section, we quotient the gauge group of the first theory by the diagonal $U(1)$ subgroup and drop the decoupled free hypermultiplet in the second theory theory. The partition function of the latter theory is
$$
{\tilde Z}_n(\eta,\omega_1,\ldots,\omega_n)=\int d\sigma \frac{e^{2\pi i \sigma\eta}}{\prod_i \cosh\pi(\sigma+\omega_i)}
$$
where $\omega_i$ is the mass parameter of the $i^{\rm th}$ hypermultiplet and $\eta$ is the FI parameter. We substitute
$$
\frac{1}{\cosh\pi (\sigma+\omega_i)}=\int d\tau_i \frac{e^{2\pi i \tau_i (\sigma+\omega_i)}}{\cosh\pi\tau_i}
$$
and integrate over $\sigma$. This gives
$$
{\tilde Z}_n(\eta,\omega_1,\ldots,\omega_n)=e^{-\frac{2\pi i}{n}\eta\sum_i \omega_i}\, \int d^n\tau\, \delta\left(\sum_i\tau_i\right) \frac{e^{2\pi i\sum_i\omega_i\tau_i}}{\prod_i \cosh\pi(\tau_i-\frac{\eta}{n})}
$$
We now define new variables
$$
\sigma_k=\sum_{i=1}^k\tau_i,\quad k=1,\ldots,n.
$$
In terms of these variables the partition function becomes
$$
{\tilde Z}_n(\eta,\omega_1,\ldots,\omega_n)=e^{-\frac{2\pi i}{n}\eta \sum_i \omega_i}\, \int d^n\sigma\, \delta(\sigma_n) \frac{e^{-2\pi i\sum_i\sigma_i (\omega_{i+1}-\omega_i)}}{\prod_i \cosh\pi(\sigma_{i+1}-\sigma_i-\frac{\eta}{n})}
$$
The latter integral is precisely the partition function of the $A_{n-1}$ quiver gauge theory with the FI parameters $\eta_i=\omega_{i+1}-\omega_i$ and a common mass $\omega=-\eta/n$ for all bifundamental hypermultiplets. The delta-function $\delta(\sigma_n)$ arises from the fact that the quotient $U(1)^n/U(1)_{diag}$ can be identified with the subgroup of $U(1)^n$ where the $n^{\rm th}$ parameter is set to zero. 

The prefactor
\begin{equation}\label{prefactor}
e^{-\frac{2\pi i}{n}\eta \sum_i \omega_i}
\end{equation}
can be attributed to an additional local term in the action which involves only the background vector multiplets \cite{Kapustin:1999ha,Witten:2003ya}. That is, when coupling a theory to background vector multiplets there is a freedom to add a gauge-invariant and supersymmetric term to the action which involves only the background vector multiplets. In the present case we couple the quiver theory to a single vector multiplet $V$ and $n$ twisted vector multiplets ${\tilde V}_1,\ldots,{\tilde V}_n$. The mass parameter $-\eta/n$ of the quiver theory arises from the $V$ background expectation value, while the FI parameter $\omega_{i+1}-\omega_i$ arises from the background expectation value of ${\tilde V}_{i+1}-{\tilde V}_i$. The prefactor (\ref{prefactor}) arises from the BF coupling of $V$ and $\sum _i {\tilde V}_i$ (with a coefficient $1$).

\section{\label{Quivers}{Nonabelian mirror pairs}}
In this section we carry out the calculation of the partition function for nonabelian mirror pairs. We first write down the matrix integral for the partition function without FI and mass parameters and find an appropriate change of variables to bring the matrix integral into a manifestly mirror-symmetric form. We then add the FI and mass parameters and compare the deformed partition functions of the mirror theories. While both in the abelian and nonabelian case the basic identity
\begin{equation}\label{sechfourier}
\int d\sigma \frac{e^{2\pi i\sigma\eta}}{\cosh\pi\sigma}=\frac{1}{\cosh\pi\eta}
\end{equation}
plays an important role, in the nonabelian case an additional trick (the Cauchy determinant formula) is needed to prove the coincidence of the partition functions of the mirror theories.

\subsection{The partition function}\label{sec:variablechange}
Consider the $\cN=4$ superconformal theory corresponding to $N$ D3 branes in the background of several D5 and NS5 branes.  As discussed above, the partition functions localizes to a matrix model whose integrand arises purely from the one-loop determinants of the vector and hypermultiplets. 

The contribution of the $U(N)$ vector multiplet arising from the $\alpha$th segment of $D3$ branes is (here the lower index labels the segment and the upper index is a Cartan subalgebra index, running from $1$ to $N$):

\[ \frac{1}{N!} \int d^N \sigma_\alpha \prod_{i<j}\sinh^2\pi({\sigma_\alpha}^i - {\sigma_\alpha}^j) \]
where:

\[ d^N \sigma_\alpha = d{\sigma_\alpha}^1 ... d{\sigma_\alpha}^N \]

denotes the integration over the eigenvalues of the zero mode of $\sigma$.

The contribution of a fundamental hypermultiplet in the $\alpha$th gauge group factor arising from a D5 brane is:

\[ \prod_i \frac{1}{\cosh\pi {\sigma_\alpha}^i} \]

Since there is one such factor for every D5 brane, we will refer to it as a D5 contribution.

The contribution of a bifundamental hypermultiplet between the $\alpha$th and $(\alpha+1)$th segment of D3 branes is:

\[ \prod_{i,j} \frac{1}{\cosh\pi({\sigma_\alpha}^i - {\sigma_{\alpha+1}}^j)} \]

Actually, given that each segment of D3 branes ends on two adjacent NS5 branes, and that the contribution of each vector multiplet is a square, it is convenient to regard the latter as product of two identical factors each of which arises from an NS5 brane. Then the NS5 contribution becomes:

\[  \frac{1}{N!} \frac{\prod_{i<j} \sinh\pi({\sigma_\alpha}^i - {\sigma_\alpha}^j) \sinh\pi({\sigma_\alpha}^i - {\sigma_{\alpha+1}}^j)}{\prod_{i,j}\cosh\pi({\sigma_\alpha}^i - {\sigma_{\alpha+1}}^j)} \]

Once we take all NS5 branes into account this will give us the correct contribution of all vector multiplets and bifundamental hypermultiplets.  The reason it is convenient to define the NS5 contribution in this way is the following identity:

\begin{equation}\label{cauchytrig}
\frac{\prod_{i<j} \sinh(x_i - x_j) \sinh(y_i - y_j)}{\prod_{i,j}\cosh(x_i - y_j)} = \sum_\rho (-1)^\rho \prod_i \frac{1}{\cosh(x_i - y_{\rho(i)})},
\end{equation}

where $\rho$ runs over all permutations of $\{1,...,N\}$.  This identity is proved in the appendix.

Using this identity, the contribution of an $NS5$ brane becomes:

\[  \frac{1}{N!}  \frac{\prod_{i<j} \sinh \pi( {\sigma_\alpha}^i - {\sigma_\alpha}^j) \sinh \pi({\sigma_{\alpha+1}}^i - {\sigma_{\alpha+1}}^j)}{\prod_{i,j} \cosh \pi ({\sigma_\alpha}^i - {\sigma_{\alpha+1}}^j)}  =  \frac{1}{N!} \sum_\rho (-1)^\rho \prod_i \frac{1}{\cosh \pi ({\sigma_\alpha}^i - {\sigma_{\alpha+1}}^{\rho(i)})} \]

\[ =  \frac{1}{N!} \sum_\rho (-1)^\rho \int d^N \tau_\alpha \prod_i \frac{e^{2 \pi i {\tau_\alpha}^i ({\sigma_\alpha}^i - {\sigma_{\alpha+1}}^{\rho(i)})}}{\cosh (\pi {\tau_\alpha}^i) } \rightarrow \mbox{NS5} \]

where we have introduced auxilliary variables ${\tau_\alpha}^i$ and used the Fourier transform identity (\ref{sechfourier}).  

For the $D5$ brane, it will be convenient to introduce a pair of auxilliary variables and write it as:

\[ \prod_i \frac{1}{\cosh (\pi \sigma^i)} = \int d^N \hsigma \prod_i \frac{\delta(\hsigma^i - \sigma^i)}{\cosh(\pi \sigma^i)} \]

\[ = \int d^N \hsigma d^N \tau \prod_i \frac{e^{2 \pi i \tau^i(\hsigma^i - \sigma^i)}}{\cosh(\pi \hsigma^i)} \rightarrow \mbox{D5} \]

Note that there is now a pair of variables ${\sigma_a}^i, {\tau_a}^i$ for each fivebrane, which are labeled by an index $a$.

Now we can consider a sequence of $D5$ and $NS5$ branes.  For concreteness, we will first take as an example the sequence $( NS5, D5, NS5 )$.  Then, using the expressions above, the corresponding partition function is given by:

\[ Z = \int  \prod_{a=1}^3 d^N \sigma_a d^N \tau_a \bigg( \frac{1}{N!} \sum_{\rho_1} (-1)^{\rho_1} \prod_i \frac{e^{2 \pi i {\tau_1}^i ({\sigma_1}^i - {\sigma_2}^{\rho_1(i)})}}{\cosh (\pi {\tau_1}^i) } \bigg) \bigg( \prod_i \frac{e^{2 \pi i {\tau_2}^i({\sigma_2}^i - {\sigma_3}^i)}}{\cosh(\pi {\sigma_2}^i)} \bigg) \times \]

\[ \times \bigg(\frac{1}{N!} \sum_{\rho_3} (-1)^{\rho_3} \prod_i \frac{e^{2 \pi i {\tau_3}^i ({\sigma_3}^i - {\sigma_1}^{\rho_3(i)})}}{\cosh (\pi {\tau_3}^i) } \bigg) \]

Note that the third factor is antisymmetric under permutations of ${\sigma_3}^i$.  This means that we can antisymmetrize over this variable in the second term, which brings the expression into a more symmetric form:\footnote{With a little thought one can see that this same idea will work in the general case provided there is at least one $NS5$ brane present.}

\[ = \int \prod_{a=1}^3 d^N \sigma_a d^N \tau_a  \bigg(\frac{1}{N!} \sum_{\rho_1} (-1)^{\rho_1} \prod_i \frac{e^{2 \pi i {\tau_1}^i ({\sigma_1}^i - {\sigma_2}^{\rho_1(i)})}}{\cosh (\pi {\tau_1}^i) } \bigg) \times \]

\[ \times \bigg(\frac{1}{N!} \sum_{\rho_2} (-1)^{\rho_2}  \prod_i \frac{e^{2 \pi i {\tau_2}^i({\sigma_2}^i - {\sigma_3}^{\rho_2(i)})}}{\cosh(\pi {\sigma_2}^i)} \bigg) \bigg(\frac{1}{N!} \sum_{\rho_3} (-1)^{\rho_3} \prod_i \frac{e^{2 \pi i {\tau_3}^i ({\sigma_3}^i - {\sigma_1}^{\rho_3(i)})}}{\cosh (\pi {\tau_3}^i) } \bigg) \]

More generally, if we label a sequence of branes by $( \alpha_1,...,\alpha_n )$, where $\alpha_a$ is either ``$D5$'' or ``$NS5$'', then we find:

\begin{equation}
\boxed{Z = \int \prod_{a=1}^{n} \frac{1}{N!} d^N \sigma_a d^N \tau_a \sum_{\rho_a} (-1)^{\rho_a} \prod_i \frac{e^{2 \pi i {\tau_a}^i ({\sigma_a}^i - {\sigma_{a+1}}^{\rho_a(i)})}}{ I_{\alpha_a}({\sigma_a}^i,{\tau_a}^i) }}
\end{equation}

where the index $a$ is to be read mod $n$, and we have defined:

\[ I_\alpha (\sigma, \tau) = \left\{ \begin{array}{cc} \cosh(\pi \sigma) & \alpha = D5 \\ \cosh(\pi \tau) & \alpha = NS5 \end{array} \right. \]

In this form, mirror symmetry is nearly manifest.  Namely, consider the numerator of the integrand:

\[ \prod_{a=1}^n \sum_{\rho_a} (-1)^{\rho_a} \prod_i e^{2 \pi i {\tau_a}^i ({\sigma_a}^i - {\sigma_{a+1}}^{\rho_a(i)})} \]

\[ = \prod_{a=1}^n \sum_{\rho_a} (-1)^{\rho_a} \prod_i e^{2 \pi i {\sigma_a}^i ({\tau_a}^i - {\tau_{a-1}}^{{\rho_{a-1}}^{-1}(i)})} \]

which, after relabeling variables $\tau_a \rightarrow - \tau_{a+1}$, gives:

\[ \rightarrow \prod_{a=1}^n \sum_{\rho_a} (-1)^{\rho_a} \prod_i e^{2 \pi i {\sigma_a}^i ({\tau_a}^i - {\tau_{a+1}}^{\rho_a(i)})} \]

Thus we see there is a symmetry under exchanging $\sigma_a$ with $-\tau_a$ in the numerator.  Performing this exchange in the denominator gives us the matrix model for the mirror theory, ie, the one we get by exchanging the $D5$ and $NS5$ branes.

\subsection{\label{BackgroundFields} Including FI and mass parameters}

It is more informative to compare the partition functions of mirror theories deformed by FI and mass parameters. Mirror symmetry predicts that they agree if the FI and mass parameters are exchanged. Each factor $U(N)$ in the gauge group gives rise to an FI parameter, so the total number of FI parameters is equal to the number of NS5 branes. Each hypermultiplet gives rise to a mass parameter, but some of them are trivial, in the sense that they can be absorbed into a shift of the scalars $\sigma$.  True mass parameters are associated with abelian global symmetries acting on the hypermultiplets. In the case of quiver theories, each bifundamental hypermultiplet has a mass parameter, but only their sum is nontrivial in the above sense, so we may assume that all bifundamentals have the same mass. There is also a mass parameter for each fundamental hypermultiplet, but only their differences are nontrivial.  Thus the total number of mass parameters is equal to the number of D5 branes. Since mirror symmetry exchanges NS5 and D5 branes, this is consistent with the proposal that mirror symmetry exchanges FI and mass parameters.

To check the mirror symmetry prediction quantitatively, we now include the FI and mass parameters in the computation of the partition functions.  As described above, an FI term $\xi$ in the gauge group corresponding to the variable $\sigma$ introduces a factor:

\[ e^{2 \pi i \xi \sum_i \sigma^i} \]

Meanwhile, a mass $m$ for a fundamental modifies its contribution to:

\[ \prod_i \frac{1}{\cosh \pi (\sigma^i + m)} \]

\noindent and similarly for a bifundamental.  Thus we start by considering the modified D5 brane contribution:
 
\[ \prod_i \frac{e^{2 \pi i \eta_a {\sigma_{a+1}}^i}}{\cosh \pi ({\sigma_{a+1}}^i + \omega_a)}  \]

In terms of the auxilliary $\tau$ variables, this can be written as:

\[ \int d^N \sigma_a d^N \tau_a \prod_i \frac{e^{2 \pi i {\tau_a}^i ({\sigma_a}^i - {\sigma_{a+1}}^i)}}{\cosh \pi ({\sigma_a}^i + \omega_a)} e^{2 \pi i \eta_a {\sigma_a}^i} \]

Mirror symmetry then tells us we should consider the following modified NS5 brane contribution:

\[ \frac{1}{N!} \int d^N \sigma_a d^N \tau_a \sum_\rho (-1)^\rho \prod_i \frac{e^{2 \pi i {\tau_a}^i ({\sigma_a}^i - {\sigma_{a+1}}^{\rho(i)})}}{\cosh \pi ({\tau_a}^i + \eta_a)} e^{2 \pi i \omega_a {\tau_a}^i} \]

To see what this corresponds to in the original matrix model, we need to integrate out the auxilliary $\tau$ variables.  We find:

\[ = \frac{1}{N!} \int d^N \sigma_a d^N \tau_a \sum_\rho (-1)^\rho \prod_i \frac{e^{2 \pi i ({\tau_a}^i-\eta_a) ({\sigma_a}^i - {\sigma_{a+1}}^{\rho(i)})}}{\cosh (\pi {\tau_a}^i)} e^{2 \pi i \omega_a ({\tau_a}^i - \eta_a)}\]

\[ = \frac{1}{N!} e^{-2\pi i \eta_a \omega_a} \int d^N \sigma_a \sum_\rho (-1)^\rho \prod_i \frac{e^{-2 \pi i \eta_a ({\sigma_a}^i - {\sigma_{a+1}}^{\rho(i)})}}{\cosh \pi ({\sigma_a}^i - {\sigma_{a+1}}^{\rho(i)} + \omega_a)} \]

\[ =  \frac{1}{N!} e^{-2\pi i \eta_a \omega_a} \int d^N \sigma \frac{\prod_{i<j} \sinh \pi ({\sigma_a}^i - {\sigma_a}^j)\sinh \pi ({\sigma_{a+1}}^i - {\sigma_{a+1}}^j) }{\prod_{i,j} \cosh \pi ({\sigma_a}^i-{\sigma_{a+1}}^j + \omega_a )} e^{2 \pi i \eta_a \sum_i ({\sigma_{a+1}}^i - {\sigma_a}^i)}  \]

which is the contribution of an NS5 brane with the corresponding bifundamental having a mass $\omega_a$, and FI terms $\eta_a$ and $-\eta_a$ respectively in the two adjacent gauge groups.  There is also an overall field independent phase, related to (\ref{prefactor}), which we will ignore.

To summarize, in terms of the parameters $\eta_a$ and $\omega_a$ that we have introduced for each fivebrane, the total FI term in the $\alpha$th gauge group factor is:

\[ \xi_\alpha = \eta_{\alpha-1} - \eta_\alpha + \sum_{a_\alpha} \eta_{a_\alpha} \]

where the first two terms come from the two NS5 branes bounding the section of D3 branes corresponding to this factor, and $a_\alpha$ runs over the D5 branes lying on this section.  Meanwhile the masses are assigned as:

\[ m^{bif}_\alpha = \omega_\alpha \]
\[ m^{fun}_{a_\alpha} = \omega_{a_\alpha} \]

Since mirror symmetry in the matrix model corresponds to the exchange of $\sigma$ with $-\tau$, we see that the parameters map acoording to $\omega \leftrightarrow -\eta$.  In terms of the mass and FI parameters, we have:\footnote{Here $\beta$ runs over gauge group factors in the mirror theory, and so also labels D5 branes in the original theory.  Similarly, $a_{\beta}$ lables NS5 branes in the original theory.  Also, we have put primes on parameters corresponding to the mirror theory.}

\[ \xi_\beta' = m^{fun}_\beta - m^{fun}_{\beta-1} - \sum_{a_\beta} m^{bif}_{a_\beta} \]

This equation (along with its dual, obtained by exchaninging the roles of the original and mirror theories) contains all the information about how the parameters map.  For example, if we define $\xi_{diag} = \sum_\alpha \xi_\alpha$, we see:

\[ \sum_\beta {m^{bif}}_\beta' = - \xi_{diag} \]

As described above, this sum contains all physical information about the masses of the bifundamentals.  One can also obtain a formula for the masses of fundamentals of the mirror theory, as we illustrate with an example.

Consider the non-abelian version of the $A_{n-1}$ mirror pair.  The first theory has gauge group $U(N)$, with $1$ adjoint hypermultiplet and $n$ fundamentals.  We can deform this by giving the masses $m_i$ to the fundamentals, $m$ to the adjoint, and an FI term $\xi$.  The mirror theory has gauge group $U(N)^n$, with a bifundamental between each pair of adjacent gauge groups, and a fundamental in one of them, say the $n$th.  We can deform this with FI terms $\xi_i'$, masses $m_i'$ for the bifundamentals, and a mass $m'$ for the fundamental.

Then from what we found above, these parameters should map as:

\[ \xi_i' = m_{i+1} - m_i \;\;\; i<n \]

\[ \xi_n' = m_1 - m_n - m \]

Comparing with equation $3.1$ in \cite{deBoer:1996mp}, we see this is the expected mapping.

\section{\label{ABJM}$\mathcal{N}=8$ super-Yang-Mills theory and the ABJM theory}

\subsection{ABJM, $\cN=8$ SYM, and its mirror}

The ABJM theory \cite{Aharony:2008ug} is an $\cN=6$ superconformal gauge theory in three dimensions. It has gauge group $G\times G$ where $G$ is either $U(N)$ or $SU(N)$. In this paper we will be interested in the case $G=U(N)$. One way to construct this theory is to consider $\cN=2$ super-Chern-Simons theory with gauge group $U(N)\times U(N)$ and four chiral multiplets: two in the bifundamental $({\bf N},\bar{\bf N})$ representation and two in the complex-conjugate representation $(\bar{\bf N},{\bf N})$. To make the theory $\cN=6$ supersymmetric one has to take the Chern-Simons levels of the two $U(N)$ factors to be opposite, $k=-{\tilde k}$, and add a certain quartic superpotential.  It has been conjectured in \cite{Aharony:2008ug} that this theory describes $N$ M2 branes on an orbifold $\CC^4/\ZZ_k\times \RR^3$. In particular, for $k=1$ the theory must describe $N$ M2 branes in flat space-time. On the other hand, it is believed that the theory of $N$ M2 branes in flat space-time can be described as the low-energy limit of $\cN=8$ super-Yang-Mills theory with gauge group $U(N)$. Hence the $k=1$ ABJM theory must be isomorphic to the low-energy limit of the $\cN=8$ super-Yang-Mills theory. 

In this section we test this conjecture by computing the partition functions of both theories on a three-sphere as a function of mass and FI deformations and comparing them. It is straightforward to write down the partition function of the ABJM theory for arbitrary $k$ \cite{Kapustin:2009kz}. On the other hand, as explained in section \ref{sec:IR}, the localization method does not apply directly to the low-energy limit of $\cN=8$ SYM theory because the superconformal R-symmetry which determines the dimensions of the fields is not manifest in the UV description of the theory. We circumvent this difficulty by replacing the $\cN=8$ SYM with $\cN=4$ theory with an adjoint and a fundamental hypermultiplet. These two theories are isomorphic in the low-energy limit, and this isomorphism may be regarded as a special case of mirror symmetry. Indeed, consider a Hanany-Witten type brane configuration with $N$ D3-branes and a single NS5 brane \cite{HW}. Such a brane configuration is described by the $U(N)$ $\cN=8$ SYM theory. Its mirror is obtained by replacing the NS5 brane with a D5 brane and is described by the $U(N)$ $\cN=4$ SYM theory coupled to a single fundamental hypermultiplet and a single adjoint hypermultiplet.

The partition function of the latter theory in the low-energy limit is given by
$$
Z_{SYM}(\eta,\omega)=\frac{1}{N!}\int d^N\sigma \frac{\prod_{i<j} \sinh^2(\pi(\sigma_i-\sigma_j))e^{2\pi i\eta\sum_i \sigma_i}}{\prod_{i,j}\cosh(\pi(\sigma_i-\sigma_j+\omega))\prod_i \cosh(\pi\sigma_i)}
$$
Here we introduced two deformation parameters: the mass parameter $\omega$ for the adjoint hypermultiplet and the FI parameter $\eta$. The mass parameter for the fundamental hypermultiplet is not physical and can be removed by a shift of the field $\sigma$.\footnote{Note that without the contribution of the fundamental hypermultiplet the partition function would not converge. This does not mean that the partition function of the low-energy limit of the $\cN=8$ SYM on a three-sphere is infinite. Rather, it means that a naive application of the localization method to $\cN=8$ SYM gives the wrong (divergent) result.} This function is actually symmetric with respect to the exchange of $\eta$ and $\omega$. To see this we first use the identity (\ref{cauchytrig}) to rewrite the partition function in the form
\begin{equation}\label{sympartition}
Z_{SYM}(\eta,\omega)=\sum_\rho (-1)^{\rho} \frac{1}{N!}\int d^N\sigma \frac{e^{2\pi i\eta\sum_i \sigma_i}}{\cosh(\pi\sigma_i)\cosh(\pi(\sigma_i-\sigma_{\rho(i)}+\omega))}
\end{equation}
Then we apply the identity (\ref{sechfourier}) to write the hyperbolic secants involving $\omega$ as their own Fourier transforms:
$$
Z_{SYM}(\eta,\omega)=\sum_\rho (-1)^{\rho} \frac{1}{N!}\int d^N\sigma d^N\tau \frac{e^{2\pi i(\eta\sum_i \sigma_i+\sum_i\tau_i(\sigma_i-\sigma_{\rho(i)}+\omega))}}{\prod_i \cosh(\pi\sigma_i)\cosh(\pi\tau_i)}
$$
Now it is easy to see that exchanging $\eta$ and $\omega$ is equivalent to exchanging $\sigma$ and $\tau$ variables. This symmetry can be explained if we use another Hanany-Witten-type realization of this theory involving $N$ D3 branes, one NS5 brane and one D5 brane. This brane configuration is self-mirror, so the theory is self-mirror, and its partition function must be symmetric under the exchange of FI and mass parameters.

Since our original goal was to understand the infrared limit of $\cN=8$ SYM theory, one might inquire about the interpretation of the deformation parameters $\omega$ and $\eta$ from the point of view of $\cN=8$ SYM. The FI parameter $\eta$ corresponds to the mass of the adjoint hypermultiplet. Indeed, consider the brane construction of the $\cN=4$ SYM with one adjoint and one fundamental hypermultiplet which involves $N$ D3 branes extended in the directions $x^0,x^1,x^2,x^3$, with $x^3$ periodic, and a single D5 brane extended in the directions $x^7,x^8,x^9$. The FI parameter corresponds to a deformation of $\RR^9\times S^1$ into an affine bundle over $S^1$ with fiber $\RR^9$ \cite{Witten:4d}. This affine bundle is defined as a quotient of $\RR^{10}$ by the following $\ZZ$-action:
$$
(x^0,x^1,x^2,x^3,x^4,x^5,x^6,x^7,x^8,x^9)\mapsto (x^0,x^1,x^2,x^3+2\pi n ,x^4,x^5,x^6,x^7+n \eta,x^8,x^9),\quad n\in\ZZ.
$$
This deformation causes the D3 branes to break at the location of the D5 brane, thereby lifting the Coulomb branch. It also causes Abrikosov-Nielsen-Olsen vortices represented by D1 branes to be massive. The mirror of this deformation causes the D3 branes to break at the location of the NS5 brane, thereby lifting the Higgs branch. It also gives a mass to the adjoint hypermultiplet represented by open strings stretching across the NS5-brane. 

On the other hand, the mirror of the mass deformation $\omega$ cannot be easily identified. The approach via brane configurations fails because this deformation is not visible in the brane construction involving a single D5 brane. It is visible in the alternative brane configuration involving one D5 and one NS5 brane, where it corresponds to a deformation of $\RR^9\times S^1$ into an affine bundle over $S^1$ similar to the one above, but with coordinates $x^7,x^8,x^9$ and $x^4,x^5,x^6$ exchanged. However this theory is self-mirror and the relationship of its infrared limit with that of $\cN=8$ SYM is highly nontrivial. The alternative brane configuration makes it clear that $\omega$ and $\eta$ are related by an element of the R-symmetry group $Spin(6)\subset Spin(8)$ which rotates the directions $x^4,x^5,x^6$ into $x^7,x^8,x^9$. However, this R-symmetry is not visible in the UV description which has only $SU(2)_R\times SU(2)_N\times SU(2)_F$ symmetry, where $SU(2)_F$ is the flavor symmetry of the adjoint hypermultiplet. 

\subsection{A comparison of the partition functions}

The partition function of the ABJM theory is
$$
Z_{ABJM}(\eta,\omega)=\frac{1}{(N!)^2}\int d^N\sigma d^N\tsigma \frac{\prod_{i<j}\sinh^2(\pi (\sigma_i-\sigma_j))\sinh^2(\pi(\tsigma_i-\tsigma_j))e^{2\pi i\zeta\sum_i(\sigma_i+\tsigma_i)+\pi i \sum_i(\sigma_i^2-\tsigma_i^2)}}{\prod_{i,j} \cosh(\pi(\sigma_i-\tsigma_j+\xi))\cosh(\pi(\sigma_i-\tsigma_j-\xi))}
$$
Here we deformed the theory by mass terms for the chiral multiplets and the FI parameter for the diagonal $U(1)$ subgroup. We assumed that the sum of the two masses is zero because it can be made zero by shifting $\sigma_i$ and $\tsigma_i$. Using the freedom to make shifts of $\sigma_i$ and $\tsigma_i$  one can also see that the difference of the FI parameters for the two $U(N)$ vector multiplets can be made zero.

Next we apply to $Z_{ABJM}$ the same kind of transformations as in section \ref{Quivers}. First we apply the identity (\ref{cauchytrig}) to write
$$
Z_{ABJM}=\sum_{\rho,\rho'}(-1)^{\rho+\rho'} \frac{1}{(N!)^2}\int d^N\sigma d^N\tsigma \prod_{i}\frac{e^{2\pi i\zeta(\sigma_i+\tsigma_i)+\pi i (\sigma_i^2-\tsigma_i^2)}}{\cosh(\pi(\sigma_i-\tsigma_{\rho(i)}+\xi))\cosh(\pi(\sigma_i-\tsigma_{\rho'(i)}-\xi))}
$$
Then we note that the integral really depends only on the composition  $\rho'\circ\rho^{-1}$, so instead of summing over both $\rho$ and $\rho'$ we may take $\rho$ to be the trivial permutation and multiply the result by $N!$. Therefore we get
$$
Z_{ABJM}=\sum_{\rho}(-1)^{\rho} \frac{1}{N!}\int d^N\sigma d^N\tsigma \prod_{i}\frac{e^{2\pi i\zeta(\sigma_i+\tsigma_i)+\pi i (\sigma_i^2-\tsigma_i^2)}}{\cosh(\pi(\sigma_i-\tsigma_i+\xi))\cosh(\pi(\sigma_i-\tsigma_{\rho(i)}-\xi))}
$$
After this we use the identity (\ref{sechfourier}) to write hyperbolic secants as their own Fourier transforms thereby introducing $2N$ integration variables $\tau_i$ and $\tau'_i$. The integral over $\sigma_i$ and $\tsigma_i$ becomes Gaussian. Computing this Gaussian integral results in
$$
Z_{ABJM}=\sum_\rho (-1)^\rho \frac{1}{N!}\int d^N\tau d^N\tau' \frac{e^{-2\pi i\sum_i\tau_i(\tau'_i-\tau'_{\rho(i)}-\xi+2\zeta)-2\pi i\sum_i \tau'_i (\xi+2\zeta)}}{\prod_i \cosh(\pi\tau_i)\cosh(\pi\tau'_i)}.
$$
Now we can perform the integral over the variables $\tau_i$ using the identity (\ref{sechfourier}). After renaming $\tau'_i\ra\tau_i$, we get
$$
Z_{ABJM}=\sum_\rho (-1)^\rho \frac{1}{N!}\int d^N\tau \frac{e^{-2\pi i(\xi+2\zeta)\sum_i \tau_i}}{\prod_i \cosh(\pi\tau_i)\cosh(\pi(\tau_i-\tau_{\rho(i)}-\xi+2\zeta))}
$$
Comparing with the partition function of the SYM theory (\ref{sympartition}) we see that the partition functions agree provided that we identify
$$
\eta=\xi+2\zeta,\quad \omega=\xi-2\zeta.
$$
Note that this duality does not merely exchange the FI and mass parameters, as was the case with mirror symmetry.

We close by mentioning that the duality between ABJM and $\mathcal{N}=8$ SYM is a special case of a more general duality \cite{Jensen:2009xh} which extends $3D$ mirror symmetry by, in addition to $NS5$ and $D5$ branes, having $(1,1)$ branes intersect the $D3$ branes.  Although we do not prove it in this paper, it can be shown that the above proof for ABJM generalizes to this wider class of dualities.

\appendix
\section{An identity for hyperbolic functions}\label{Identity}
In this section we prove the identity:

\[ \frac{\prod_{i<j} \sinh(x_i - x_j) \sinh(y_i - y_j)}{\prod_{i,j}\cosh(x_i - y_j)} = \sum_\rho (-1)^\rho \prod_i \frac{1}{\cosh(x_i - y_{\rho(i)})} \]
Multiplying both sides by $e^{-\sum_i (x_i + y_i)}$, and defining $u_i = e^{2x_i}, v_i=e^{2 y_i}$, this becomes equivalent to:

\[ \frac{\prod_{i<j} (u_i - u_j) (v_i-v_j)}{\prod_{i,j}(u_i+v_j)} = \sum_\rho (-1)^\rho \prod_i \frac{1}{u_i + v_{\rho(i)}} \]
This identity is called the Cauchy determinant formula; since it is not widely known, we sketch a proof below.

Multiplying both sides by the denominator on the LHS we get

\[ \prod_{i<j} (u_i - u_j) (v_i-v_j) = \prod_{i,j}(u_i+v_j) \sum_\rho (-1)^\rho \prod_i \frac{1}{u_i + v_{\rho(i)}} \]
Note that both sides are completely antisymmetric under  permutations of the $u_i$, or separate permutations of the $v_i$.  The RHS can also be written as follows:

\[  \sum_\rho (-1)^\rho \prod_i \prod_{j \neq \rho(i)} ( u_i + v_j) \]
In this form we see both sides are polynomials of total degree $N(N-1)$.  We claim this property of a function, being a polynomial in $u_i,v_i$ of total degree $N(N-1)$ which is totally antisymmetric under separate permutations of the $u_i$ and the $v_i$, completely determines the function up to an overall constant.

To see this, note that in each term of the polynomial, no two $u_i$ can appear raised to the same power, since when we antisymmetrize under exchange of these two variables such terms vanish.  Thus they must all appear raised to different powers, and so the lowest degree term has the form

\[ {u_1}^0 {u_2}^1 ... {u_N}^{N-1} \]
or some permutation thereof.  The same is true of the $v_i$.  But note that this polynomial already has degree $N(N-1)$.  Thus all terms in the polynomial are permutations of this basic monomial:

\[ {u_1}^0 {u_2}^1 ... {u_N}^{N-1} {v_1}^0 {v_2}^1 ... {v_N}^{N-1} \]
Moreover, the relative coefficient between these terms is fixed by antisymmetry, so that the function must have the form:

\[ C \sum_{\pi_1,\pi_2} (-1)^{\pi_1 + \pi_2} {u_{\pi_1(1)}}^0 {u_{\pi_1(2)}}^1 ... {u_{\pi_1(N)}}^{N-1} {v_{\pi_2(1)}}^0 {v_{\pi_2(2)}}^1 ... {v_{\pi_2(N)}}^{N-1} \]
This proves the identity up to a multiplicative constant.  It can be checked that this constant is actually one, but since this is unimportant for the proof of the duality we omit this check.

\bibliographystyle{jhep}

\end{document}